\documentclass[a4paper,11pt]{article}
\usepackage{graphicx,amssymb,amsmath}
\usepackage{epsfig}

\begin{document}

%\begin{runninglinenumbers}
%\renewcommand\linenumberfont{\normalfont\small\bfseries}
%\setlength\linenumbersep{1cm}

\begin{center}
{\bf \Large The Outer Tracker Detector of the HERA-B Experiment}

{\bf \Large Part I: Detector}

\vspace{10mm}
%%\today
%%\vspace{10mm}

\begin{bf}
%
% TITLE
%
HERA-B Outer Tracker Group
\end{bf}

\vspace{10mm}               % Makes a line with 10mm blank 

H.Albrecht${}^h$, Th.S.Bauer${}^{a,q}$, M.Beck${}^p$,
K.Berkhan${}^r$, G.Bohm${}^r$, M.Bruinsma${}^{a,q}$, T.Buran${}^o$,
M.Capeans${}^h$, B.X.Chen${}^d$, H.Deckers${}^e$, A.Donat${}^r$, X.Dong${}^c$,
R.Eckmann${}^b$, D.Emelianov${}^h$, G.Evgrafov${}^{h,s}$,
I.Golutvin${}^g$, U.Harder${}^r$, M.Hohlmann${}^h$, K.H\"opfner${}^h$,
W.Hulsbergen${}^a$, Y.Jia${}^c$, C.Jiang${}^c$, H.Kapitza${}^f$,
S.Karabekyan${}^{p,u}$, Z.Ke${}^c$, Y.Kiryushin${}^g$,
H.Kolanoski${}^e$, 
D.Kr\"ucker${}^e$, A.Lanyov${}^g$, Y.Q.Liu${}^d$, T.Lohse${}^e$,
H.L\"udecke${}^r$, R.Mankel${}^e$, G.Medin${}^e$, E.Michel${}^h$,
A.Moshkin${}^g$, J.Ni${}^d$, S.Nowak${}^r$, M.Ouchrif${}^{a,q}$,
C.Padilla${}^h$, R.Pernack${}^p$, A.Petrukhin${}^{h,n}$, 
D.Pose${}^{g,j}$, B.Schmidt${}^h$, W.Schmidt-Parzefall${}^i$,
A.Schreiner${}^r$, H.Schr\"oder${}^{h,p}$, U.Schwanke${}^r$,
A.S.Schwarz${}^h$, I.Siccama${}^h$, K.Smirnov${}^r$, S.Solunin${}^g$, 
S.Somov${}^h$, 
V.Souvorov${}^r$, A.Spiridonov${}^{r,n}$, 
C.Stegmann${}^{r,e}$, O.Steinkamp${}^a$,
N.Tesch${}^h$, I.Tsakov${}^{h,t}$, U.Uwer${}^{e,j}$, S.Vassiliev${}^g$,
D.Vishnevsky${}^g$, I.Vukotic${}^e$, M.Walter${}^r$, J.J.Wang${}^d$,
Y.M.Wang${}^d$, R.Wurth${}^h$, J.Yang${}^d$, Z.Zheng${}^c$,
Z.Zhu${}^c$, R.Zimmermann${}^p$ \\
\vspace{10mm}
\footnotesize${}^a$NIKHEF, Kruislaan 409, PO Box 41882, 1009 DB Amsterdam, 
Netherlands$^1$\\
\footnotesize${}^b$University of Texas at Austin, Department of Physics, 
RLM 5.208, Austin TX 78712-1081, USA$^2$ \\
\footnotesize${}^c$Institute of High Energy Physics, Beijing 100039, China\\
\footnotesize${}^d$Institute of Engineering Physics, Tsinghua University, 
Beijing 100084 , P.R. China\\
\footnotesize${}^e$Institut f\"ur Physik, Humboldt-Universit\"at zu Berlin, 
D-12489 Berlin, Germany$^3$\\ 
\footnotesize${}^f$Institut f\"ur Physik, Universit\"at Dortmund, D-44221 
Dortmund, Germany$^3$\\ 
\footnotesize${}^g$Joint Institute for Nuclear Research, Dubna, RU-141980, 
Russia\\
\footnotesize${}^h$DESY, Notkestra{\ss}e 85, D-22607 Hamburg, Germany\\
\footnotesize${}^i$Institut f\"ur Experimentalphysik, Universit\"at Hamburg, 
D-22761 Hamburg, Germany$^3$\\
\footnotesize${}^j$Physikalisches Institut, Universit\"at Heidelberg, 
D-69120 Heidelberg, Germany$^3$\\
\footnotesize${}^n$Institute of Theoretical and Experimental Physics, 
117259 Moscow, Russia\\
\footnotesize${}^o$Institute of Physics, University of Oslo, 
Norway$^4$\\
\footnotesize${}^p$Fachbereich Physik, Universit\"at Rostock, 
D-18051 Rostock, Germany$^3$\\
\footnotesize${}^q$Universiteit Utrecht/NIKHEF, 3584 CB Utrecht, 
The Netherlands$^1$\\
\footnotesize${}^r$DESY, Platanenallee 6, D-15738 Zeuthen, 
Germany\\

\footnotesize${}^s$visitor from Moscow Physical Engineering Institute, 
115409 Moscow, Russia\\
\footnotesize${}^t$visitor from Institute for Nuclear Research, INRNE-BAS, 
Sofia, Bulgaria \\
\footnotesize${}^u$visitor from Yerevan Physics Institute, Yerevan, Armenia \\

\vspace{5mm}
%} %supported by: %\vspace{5mm}
\footnotesize$^{1}$supported by the Foundation for Fundamental Research on
Matter (FOM), 3502 GA Utrecht, The Netherlands\\
\footnotesize$^{2}$supported by the U.S. Department of Energy (DOE)\\
\footnotesize$^{3}$Supported by Bundesministerium f\"ur Bildung und Forschung, 
FRG, under contract numbers \\ 
\footnotesize$^{ }$ 05-7BU35I, 05-7D055P, 05 HB1KHA, 05 HB1HRA, 05 HB9HRA, 05
7HD15I, 
05 7HH25I \\ 
\footnotesize$^{4}$Supported by the Norwegian Research Council\\

\date{}

\vspace{10mm}

\begin{abstract} 

The HERA-B Outer Tracker is a large system of planar drift chambers
with about 113\,000 read-out channels. Its inner part has been designed 
to be exposed to a particle flux of up to $2\cdot 10^5$\,cm$^{-2}$s$^{-1}$,
thus coping with conditions similar to those expected for future hadron
collider experiments. 13 superlayers, each consisting of two individual
chambers, have been assembled and installed in the experiment. The 
stereo layers inside each chamber are composed of honeycomb drift tube 
modules with 5 and 10\,mm diameter cells. Chamber aging is prevented 
by coating the cathode foils with thin layers of copper and gold,
together with a proper drift gas choice.
Longitudinal wire segmentation is used to limit the occupancy in the
most irradiated detector regions to about 20\,\%. The production of 
978 modules was distributed among six different laboratories and took 15 
months. For all materials in the fiducial region of the detector good
compromises of stability versus thickness were found. 
A closed-loop gas system supplies the Ar/CF$_4$/CO$_2$ gas mixture to 
all chambers. The successful operation of the HERA-B Outer Tracker
shows that a large tracker can be efficiently built and safely operated
under huge radiation load at a hadron collider.

\end{abstract} 

%%% Local Variables: 
%%% mode: latex
%%% TeX-master: "main"
%%% End: 

\end{center}

%%% Local Variables: 
%%% mode: latex
%%% TeX-master: t
%%% End: 

\section{Introduction}

HERA-B is a fixed target experiment using the 920\,GeV proton beam of the HERA 
electron-proton collider \cite{proposal,tdr}. Proton-nucleus
interactions are produced using an internal wire target in the 
halo of the proton beam which allows for concurrent operation
with the $ep$ collider experiments.  
The design of the experiment was driven primarily by the physics goal of 
studying CP violation in the decay $B^0 \rightarrow J/\psi K^0_S$,
requiring high event rates and particle flux, efficient 
particle identification and triggering.

\begin{figure}
   \epsfig{file=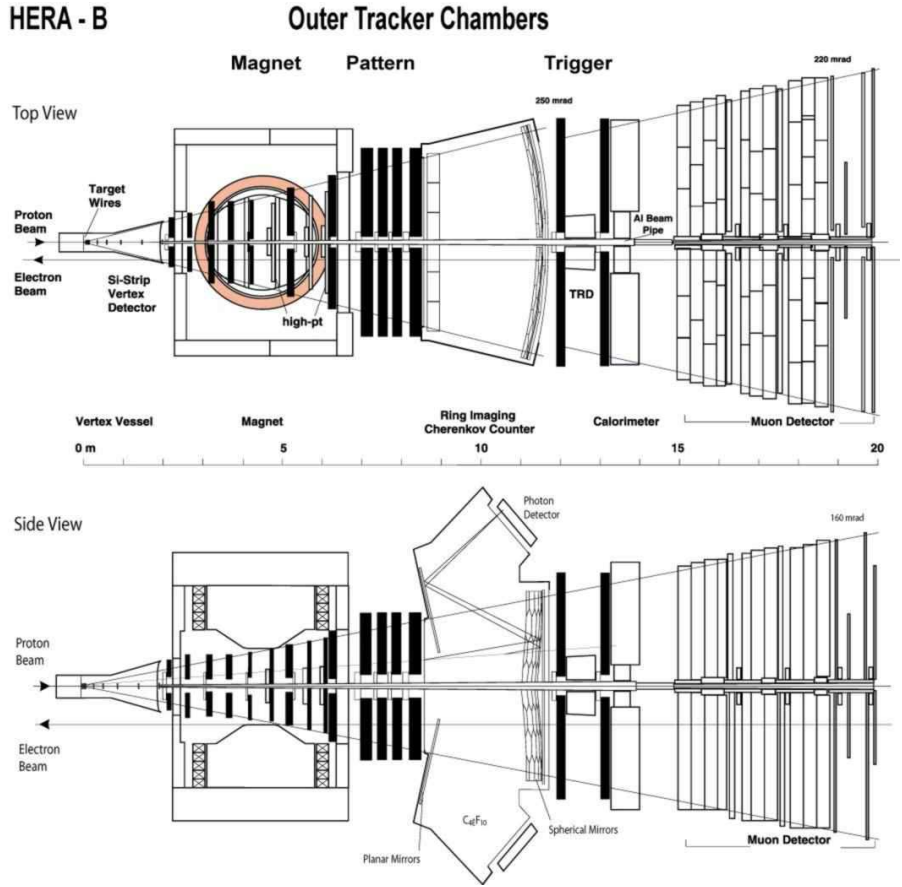,width=\linewidth}
   \caption{Schematic top and side views of the HERA-B experiment.
            The Outer Tracker chambers are indicated as black bars.
            The Inner Tracker modules are attached to them near the 
            beam pipe (small white boxes).}
   \label{fig:herab}
\end{figure}

The HERA-B spectrometer, schematically shown in Fig.\ \ref{fig:herab},
covers in the bending plane an angular range from 10\,mrad to 220\,mrad, 
corresponding to about 90\,$\%$ solid-angle coverage in the center-of-mass 
system. 

The main components of the spectrometer are the following: 
\begin{itemize}
\item The halo target station \cite{target} consists of two sets of 4 wires, 
  separated along the beam by 5\,cm. This allows to localize the vertices of 
  proton-nucleus interactions at very high rates of up to 40\,MHz with 4 to 5 
  interactions per proton bunch. 
  Most data were taken using carbon and tungsten targets, but for the
  study of $A$ dependencies other materials such as titanium and aluminum 
  have been used as well.
\item The vertex detector system (VDS) \cite{vds}, based on silicon 
  strip detector technology, starts at 1\,cm radius from the beam. 
  The detector consists of 8 superlayers extending over a length of  
  2\,m along the beam. This allows a stand-alone pattern 
  recognition of tracks and the reconstruction of both interaction 
  and decay vertices.
\item The normal-conducting dipole magnet for momentum analysis has a 
  field integral of 2.2\,Tm. The magnet aperture is 160\,mrad vertically
  and 220\,mrad horizontally.
\item The main tracking system consists of detector components with 
  different granularities and technologies varying with the distance 
  from the proton beam pipe. 
  To limit the occupancy of detector cells and to minimize the 
  number of channels, the tracking system is divided into an inner and an 
  outer part. For the Inner Tracker (ITR) \cite{itr} 
  microstrip gas chambers with 
  a gas electron multiplier (GEM) foil are used covering the regions between 
  about 6\,cm and 20 to 25\,cm from the beam axis. The strip pitch is
  300\,$\mu$m.
  The Outer Tracker (OTR), based on drift chamber technology, will be 
  discussed in this paper.
\item The particle identification system includes 
  the ring imaging Cherenkov counter (RICH) \cite{rich}, 
  the electromagnetic calorimeter (ECAL) \cite{ecal} and a muon 
  identification system (MUON) \cite{muon}. 
  The latter two are also used to trigger on $e^+e^-$ and 
  $\mu^+\mu^-$ pairs
  from $J/\psi$ decays as well as on leptons from semileptonic decays.
\end{itemize}

The detector is optimized to run at interaction rates up to 40\,MHz 
and to deal with up to 200 charged particle tracks every 96\,ns. This 
results in a high radiation load, comparable to conditions at LHC experiments.
To guarantee that the detector components sustain a running period 
of about 5 years, extensive aging studies for most of the components 
were performed \cite{hbaging}.

A selection of the decay mode $B \rightarrow J/\psi K^0_S$  
requires a trigger system which reduces the background by 
a factor of about $10^5$.
The HERA-B detector with its trigger and data acquisition system is 
flexible enough to access a wide range of additional physics topics
like $B$ production cross section, heavy-quark physics, $J/\psi$ and 
$\Upsilon$ production mechanisms as well as minimum bias physics. 
Detailed descriptions of the HERA-B detector and its capabilities
are given elsewhere \cite{proposal,tdr,hb2000}.

In the following sections we will discuss the design and construction 
of the Outer Tracker
with emphasis on the drift chamber modules, the assembly of the detector,
and the gas system. The OTR electronics and the detector performance
are described in separate papers 
\cite{otr_fee, perform}.

%%% Local Variables: 
%%% mode: latex
%%% TeX-master: t
%%% End: 

\section{General Design Aspects}

\subsection{Detector Requirements} 
The design of the Outer Tracker and the choice of technology is 
driven by requirements of the physics program, space constraints in the 
experimental area, cost optimization, and feasibility of large-scale 
production. 
The OTR is designed to serve the following purposes:
\begin{itemize}
\item Efficient reconstruction of charged particle tracks from
a distance of 20\,cm from the HERA proton beam up to the 
outer acceptance limit of the experiment.
\item Precise momentum measurement together with the vertex detector system
and the dipole magnet.
\item Providing fast trigger signals for the first level trigger (FLT)
in the environment of high track density.
\item Tracking inside the magnet to provide vertex and momentum information
of $K^0_S$ mesons decaying in the magnet.
\end{itemize}

These demands require an OTR detector with good 
performance in terms of rate capability, spatial resolution, and
hit and tracking efficiencies. It also requires a fast drift gas, for
which an Ar/CF$_4$/CO$_2$ mixture was chosen.
In addition, the amount of material 
in the tracking detector has to be small to minimize 
effects of bremsstrahlung, photon conversion, multiple scattering, 
and hadronic interactions. 

\subsection{Particle Rate and Occupancy}
\begin{figure}
   \epsfig{file=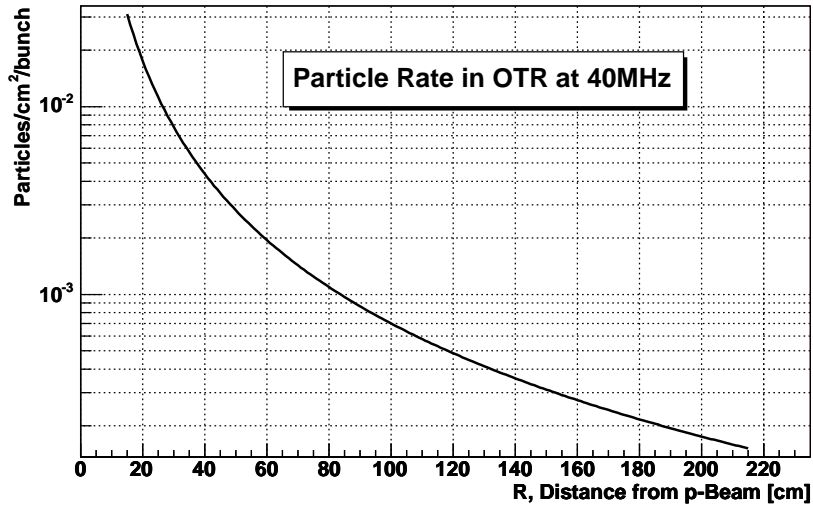,width=\linewidth}
   \caption{Rate of charged particles at 
     40\,MHz interaction rate in dependence on the distance R
     from the proton beam axis.}
   \label{fig:rate_otr}
\end{figure}

One of the most important requirements is the high rate capability of 
the OTR. 
At 40\,MHz interaction rate the detector has to deal with about 100 charged 
primary  and secondary particles per proton bunch. 
The radial dependence of the particle flux is $\approx$ 
$10^8/R^2$ particles per second. Here $R$ denotes the radial 
distance from the proton beam (Fig.\ \ref{fig:rate_otr}).

In the inner part of the OTR, at $R$ = 20\,cm, the hit occupancy 
in an event is almost 2\,$\%$ per $\textrm{cm}^{2}$ of the detector
cross section. The occupancy drops by two orders of
magnitude in the outer region at $R$ = 200\,cm. 
Efficient pattern recognition of all charged tracks and the FLT 
tracking of charged leptons demand detector occupancies of less 
than 20\,$\%$. 
For higher occupancies the rate of misreconstructed tracks 
increases and both, pattern 
recognition and tracking on the FLT level become inefficient.  
High occupancies require a high detector granularity and this 
influences strongly the choice of detector technology and read-out 
electronics. 
Compared with previous and existing experiments occupancies of 20\,$\%$
are very high. 

\subsection{\label{global}Global Outer Tracker Design}
The Outer Tracker system consists of three different parts:
 \begin{itemize}
 \item Magnet Chambers (MC) for tracking inside the magnetic field,
 \item Pattern Recognition Chambers (PC) in the field-free region between
   the magnet and the RICH for track reconstruction and for triggering,
 \item Trigger Chambers (TC) for the tracking of leptons
   in front of the ECAL and for triggering.
 \end{itemize}

\begin{table}[htb]
   \caption{OTR superlayers: Positions along the beam ($\textrm{z}_{min}$), 
     dimensions and layer structure.
     The symbols $+$ 0 $-$ denote the stereo angles of +80, 0 
     and $-$80\,mrad. The circled symbols denote double layers for the FLT.}
   \begin{center}
%%      \begin{tabular}{|c|c|c|c|c|c@{\,}c@{\,}c@{\,}c@{\,}c@{\,}c|}
         \begin{tabular}{|c|c|c|c|c|c|}
         \hline\hline
         Superlayer & $\textrm{z}_{min}$ & Thickness & Width & Height
%%         & \rule[-2mm]{0mm}{6mm} & \multicolumn{7}{c|}{Layers} \\
         & Layers \\
                    & [cm] & [cm] & [cm] & [cm] &  \\
         \hline
         MC1 & 219  & 10 & 116 &  75 & $+$ 0 $-$ \\
         MC2 & 266  & 10 & 140 &  90 & $+$ 0 $-$ \\
         MC3 & 315  & 15 & 167 & 122 & $+$ 0 $-$ \\
         MC4 & 365  & 15 & 196 & 122 & $+$ 0 $-$ \\
         MC5 & 422  &  5 & 208 & 129 & 0  \\
         MC6 & 513  & 18 & 277 & 150 & $+$ 0 $-$ \\
         MC8 & 621  & 18 & 332 & 206 & $+$ 0 $-$ \\
         \hline
         PC1 & 702  & 28 & 416 & 276 & 0 $-$ 0 $\oplus$ $\odot$ $\ominus$ \\
         PC2 & 742  & 23 & 416 & 276 & 0 $+$ 0 $-$ 0 $+$ \\
         PC3 & 778  & 23 & 416 & 276 & 0 $+$ 0 $-$ 0 $+$ \\
         PC4 & 823  & 28 & 416 & 276 & 0 $-$ 0 $\oplus$ $\odot$ $\ominus$ \\
         \hline
         TC1 & 1192 & 20 & 524 & 446 & $\oplus$ $\odot$ $\ominus$ \\
         TC2 & 1306 & 20 & 580 & 446 & $\oplus$ $\odot$ $\ominus$ \\
         \hline\hline
      \end{tabular}
   \end{center}
   \label{tab:otr_sl}
\end{table}

The track reconstruction requires a sufficient number 
of hits measured in each of the three tracker parts. 
As shown in Fig.\ \ref{fig:herab}, the Outer Tracker comprises 13 
so called superlayers (7 MC, 4 PC and 2 TC) each of which is vertically
divided into two chambers. 
Each chamber is an independent device with its own gas and 
power connections and its own read-out electronics system.

In order to allow a 3-dimensional track measurement, all superlayers 
but MC5 are combinations of three types of stereo
layers with wires at angles of 0, 80 and -80\,mrad w.r.t. the 
vertical direction. The small stereo angle helps to suppress hit 
ambiguities during pattern recognition, at the expense of an 
increased spatial resolution in the vertical direction.
Table\ \ref{tab:otr_sl} summarizes for all superlayers the position 
of the front plane, the dimensions and the layer structure. 

A viable technology for building a tracking detector with the 
specified dimensions are drift tubes, where each anode wire is located 
in the center of a cathode tube made of a thin conductive foil.
Compared to drift chambers with cathode wires, 
drift tubes offer several advantages: Due to the much lower electric
field on the cathode it is easier to avoid aging effects,
single broken wires are less dangerous for the functionality of the
whole detector, and the tubes are mechanically self-supporting. The
last fact makes it easy to group drift tubes in modules. The module
size can be chosen to best fit the requirements of detector mass
production and assembly.

Tracking in a high occupancy environment benefits from drift cells
being as small as possible, but the operational safety of the detector
sets a lower limit on the cell size. For the Outer Tracker a minimal
drift tube diameter of 5 mm is chosen. Due to the $1/r^2$ dependence of
the particle rate, tubes with double diameter (10 mm) can be used in the
outer parts of the detector. This helps to minimize the channel count.

\begin{figure}
   \epsfig{file=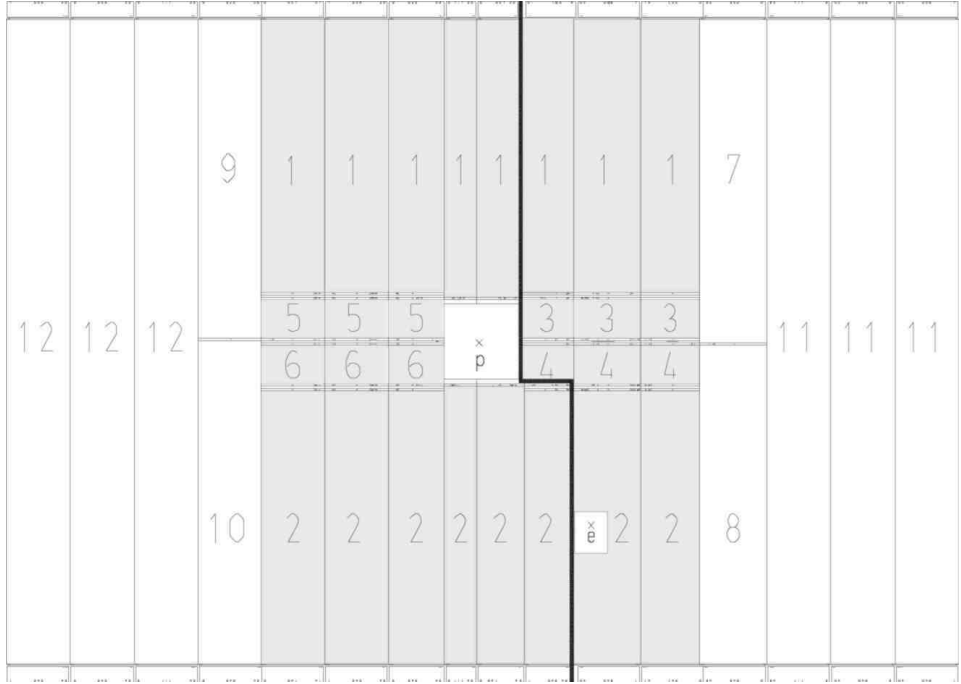,width=\linewidth}
   \caption{Segmentation of modules in an OTR detector plane. 
     Modules with 5 mm cell size (grey) are segmented in up to four 
     sectors (sector numbers 1 - 6).
     Modules with 10 mm cell size (white) have one or two sectors
     (sector numbers 7 - 12).
     Segmented modules are read out from top and bottom. 
     The thick black line indicates the mechanical separation of the 
     two chambers forming a superlayer.}
   \label{fig:cell_segm}
\end{figure}

According to Fig.\ \ref{fig:rate_otr}, there are up to 0.02 
particles/$\textrm{cm}^{2}$/bunch in the innermost area of the OTR. 
Because the cell occupancy is required to be smaller than 20\,$\%$,
the maximum cell length would be limited to 20 cm for 5 mm cells.
Since the vertical dimension of the OTR ranges from 75 to 450\,cm, 
the only solution is a longitudinal cell segmentation. 
The technical realization is described in section 3.1.
In Fig.\ \ref{fig:cell_segm} the composition of a detector plane 
from individual detector modules and the segmentation into 
different sectors is shown schematically. 
Except for the outermost sectors
11 and 12, the anode wires are separated in the middle to 
obtain two independent parts of a cell, read out from top and bottom. 
In addition, 5 mm cells are segmented into inner sectors (3 - 6) with   
20 cm long anode wires and outer sectors (1 and 2) with wire lengths 
of 25 - 205 cm, depending on the superlayer.

At the HERA-B detector, the HERA proton and electron beam 
lines are separated. The proton beam axis defines the central HERA-B 
detector axis.  
The electron beam pipe also passes through the detector.
To minimize the dead area around the beam pipes and to allow an easy
access, the superlayers are divided into two asymmetric chambers, where
the separation follows the positions of the two beam pipes
(see Fig.\ \ref{fig:cell_segm}). 

%%% Local Variables: 
%%% mode: latex
%%% TeX-master: t
%%% End: 

\section{Detector Modules}

\subsection{Design Aspects}
%%=============================

As was described in the previous section, the Outer Tracker is
assembled from modules of drift tubes with 5 or 10 mm diameter and
walls made of a light-weight conductive plastic foil. Instead of
building modules from individual tubes, a honeycomb structure 
of hexagonal drift cells is constructed layer
by layer from appropriately folded foils. The main advantage of this
technology is the possibility to assemble the anode wires into open
cells. This makes it easy to implement the longitudinal anode wire
segmentation shown in Fig.\ \ref{fig:cell_segm}.

Prototypes of drift chambers consisting of honeycomb cells with 15 mm cell
diameter have been developed for the muon 
detector of the ATLAS experiment \cite{nikhef_hcsc, aachen_hc}.
Tests using conductive soot-loaded polycarbonate foil \cite{aachen_hc,  
bayer, lonza} and an automated folding device \cite{nikhef_hcsc}  
showed that cells 
with only 5 mm diameter could be produced as well.

\begin{figure}
   \epsfig{file=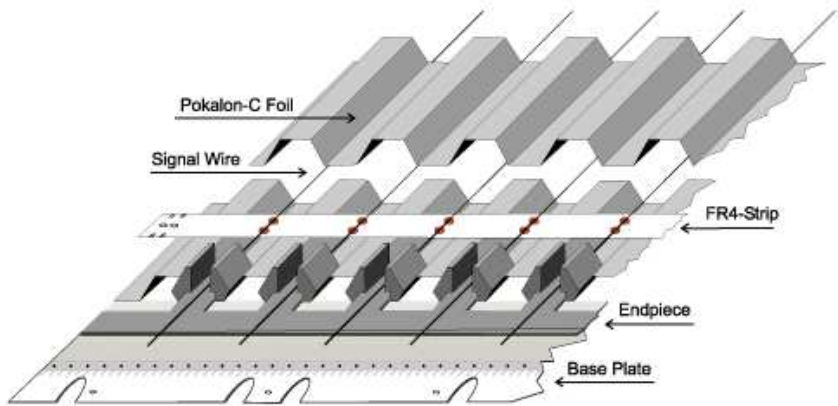,width=12cm}
   \caption{Schematical view of the module frontend composition 
     and the basic elements building up a drift cell module.}
   \label{fig:mod_prod}
\end{figure}

\begin{figure}
   \epsfig{file=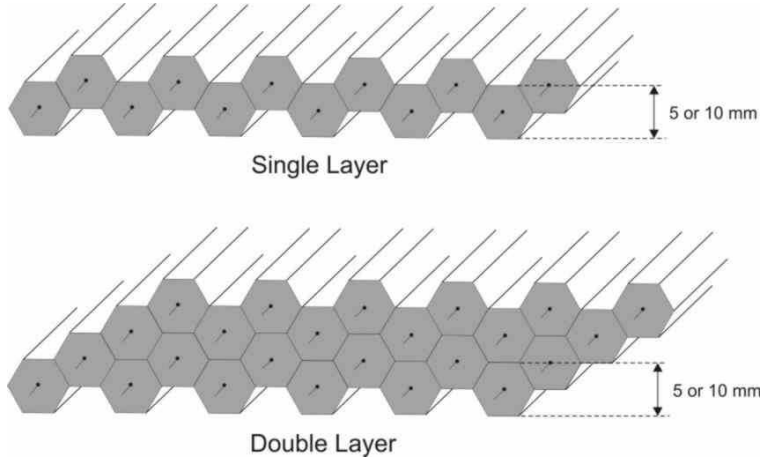,width=10cm}
   \caption{Schematical view of the single and double layer module
     cross sections.}
   \label{fig:single_doublelayer}
\end{figure}

\begin{figure}
   \epsfig{file=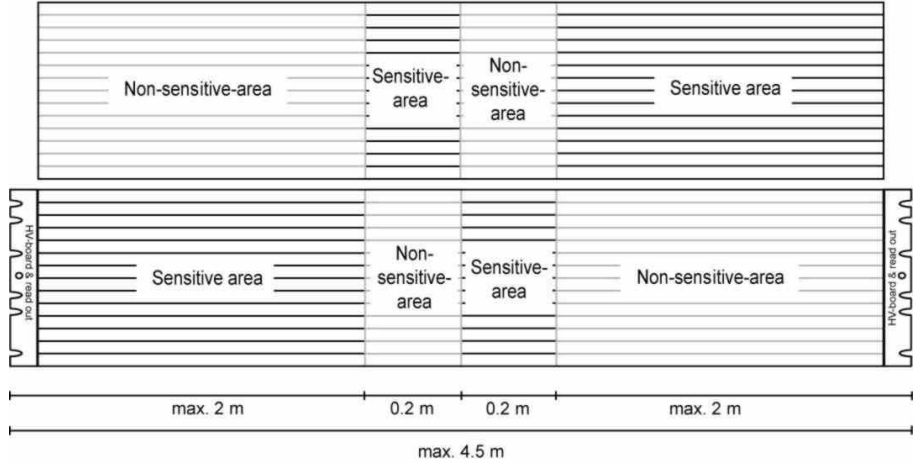,width=12cm}
   \caption{Structure of a standard 5\,mm module providing four 
     sensitive sectors. The lower single layer has one sensitive area 
     with 25\,$\mu$m anode wires in the left part which is directly 
     connected to the high voltage board. 
     The other sensitive area is in the inner right 
     part, which is read out from the right side via 75\,$\mu$m 
     thick (non-sensitive) wires. 
     The structure of the upper single layer is complementary.}
   \label{fig:segmentation}
\end{figure}

The assembly steps are indicated in Fig.\ \ref{fig:mod_prod}. 
One can see the first wired cell layer before gluing the upper foil 
on the lower one to form a so-called monolayer.
The simplest fully efficient detector plane, a so-called single
layer, consists of two staggered monolayers, as shown in 
Fig.\ \ref{fig:single_doublelayer}. 
Single layer modules are used for detector planes which are not 
implemented in the trigger. 
They have no segmentation in case of 10\,mm cells (sectors 11 and 12 in 
Fig.\ \ref{fig:cell_segm}) and for modules with 5 mm cells installed 
above and below the proton beam pipe. 
The 10\, mm modules of sectors 7 to 10 are divided into 
an upper and a lower part.
The double layer structure shown in Fig.\ \ref{fig:single_doublelayer}
is used for two classes of modules. First, it is needed to build 5\,mm 
modules with 4 sensitive sectors (labelled 1-3-4-2 and 1-5-6-2 in 
Fig.\ \ref{fig:cell_segm}).  
The details of the single layer structure for such a module are 
shown in Fig.\ \ref{fig:segmentation}, where also the longitudinal 
wire segmentation is described.
In the second class of double layer modules both single layers have 
the same structure. This creates sensitive sectors with double coverage,
where the logical 
OR of lined-up wires can be used to increase the hit efficiency. 
All OTR layers used in the FLT are of this type.

\subsection{Module Types}
%%===========================

The large-scale manufacturing of modules for superlayers of
different structure and dimensions requires a detailed planning and 
preparation. 
\begin{figure}
   \epsfig{file=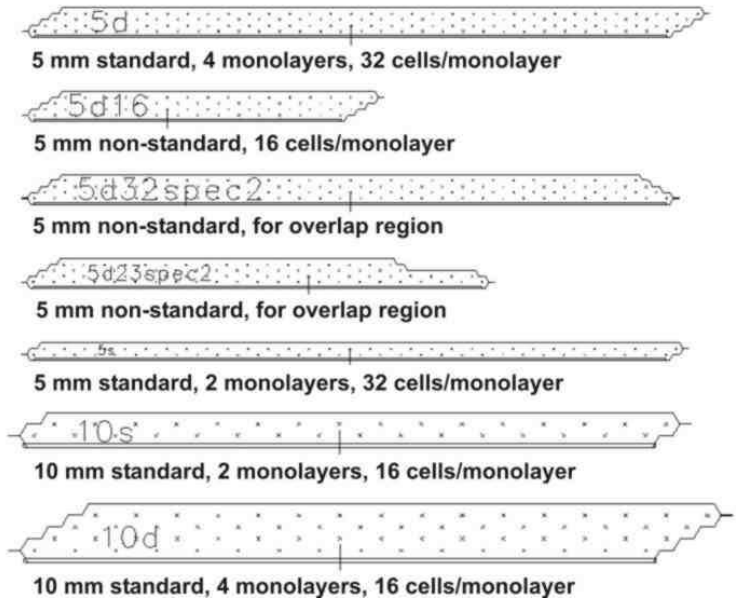,height=8cm}
   \caption{Schematics of typical module cross sections. The small 
     crosses indicate the anode wires. 
     For clarity, the honeycomb structure is not drawn.}
   \label{fig:mod_types}
\end{figure}
To minimize the number of different module types identical 
heights are chosen for the four PC and for the two TC 
superlayers (see Tab.\ \ref{tab:otr_sl}).
So-called standard modules with 5\,mm drift cells are 
defined to have 32 cells per monolayer, which is a multiple of 
the 16 channels per signal read-out board \cite {otr_fee}. 
For modules with 10\,mm cells there are 16 cells per monolayer.
This results in a module 
width of about 30\,cm and is also a reasonable size for the complicated 
and heavy 
production templates (see Section\ \ref{sec:subsectPrTools}). 
Another advantage is that even the 4.5\,m 
long TC modules weigh only 1.5\,kg and are light enough for an easy handling 
during manufacturing and for the installation into the superlayers.

\begin{figure}
   \epsfig{file=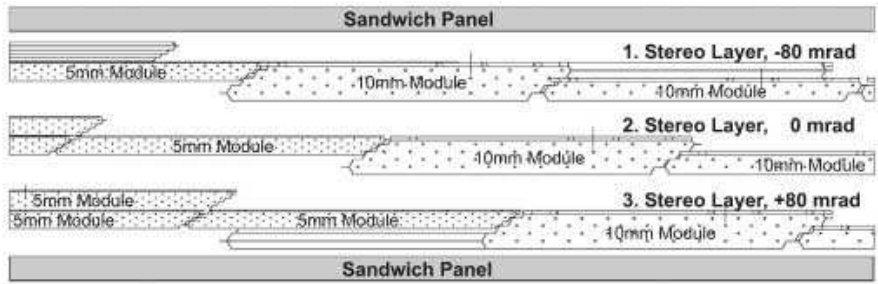,height=4cm}
   \caption{Schematics of the module overlap within a
     chamber. The picture shows a cut-out of the three
     module layers (-80\,mrad, 0\,mrad, +80\,mrad) seen from top.}
   \label{fig:mod_overl}
\end{figure}

The complete Outer Tracker contains 978 honeycomb drift tube modules.
Two thirds of these have drift cells with 5\,mm diameter. 
In total, 148 types of modules with different widths, lengths, and 
internal structures had to be produced. 
There are 25 types of standard modules, just
differing in length and/or drift cell size, which account for half the
modules in the detector.
Non-standard modules are needed mainly in the overlap of the two 
chambers of a superlayer, and around the proton and electron beam pipes.
A few examples of module cross sections are shown in 
Fig. \ref{fig:mod_types}. 
With the trapezoidal module cross section, the cells of neighbouring 
modules are overlapping. As shown in Fig.\ \ref{fig:mod_overl} for 5\,mm 
and 10\,mm modules this overlap avoids an efficiency loss. 

\subsection{Module Components}
%%============================

As shown in Fig.\ \ref{fig:mod_prod}, a module essentially consists
of cathode foils, wire support strips, anode wires in the 
sensitive part and of end-pieces and base plates on both module 
ends for the fixation. The components and materials used in
the module production will be described in more detail in the 
following. In Table\ \ref{tab:material} the product 
names, the producer and the approximate amount of all materials 
are summarized.

\begin{table}[htb]
 \caption{Materials used in the chamber production.} 
 \begin{center}
 \begin{tabular}{|p{2cm}|p{4.5cm}|p{1.5cm}|p{2.5cm}|}
 \hline\hline 
 {Component} & {Material} & {Producer} & {Amount} \\ 
 \hline\hline
 Cathode & Pokalon-C foil, 75\,$\mu$m thick & \cite{lonza} & $\sim$12000\\
  & coated with Cu (50\,nm) and Au (40\,nm) &  \cite{apvv} & 1\,m long foils
 \\\hline
 Base Plates & FR4 (glass fiber laminate, matrix epoxy resin) &
 \cite{ilfa} & $\sim$2000
 \\\hline
 End-pieces & Noryl (polyphenylen ether) & \cite{tauten} &
 $\sim$20000 \\\hline
 Wire Support Strips  & FR4 & 
 \cite{ilfa} & $\sim$24000
 \\\hline
 Glue & Stycast 1266A and Catalyst 9  & \cite{emmerson} & $\sim$40\,kg
 \\\hline
 Conductive Glue & E-solder 3025\,A+B & \cite{epoxy} & $\sim$1.5\,kg 
 \\\hline
 Signal Wire & 25\,$\mu$m gold-plated tungsten wire & \cite{calfw} &
 $\sim$220000\,m \\\hline
 Thick Wire & 75\,$\mu$m Cu + Be wire & \cite{littlfa} & 
 $\sim$150000\,m \\\hline
 Solder Tin & FLUITIN Sn60Pb & \cite{kueppers} & $\sim$10\,kg
 \\\hline
 Carbon Fiber Rods & carbon fiber composite, 2\,mm diam., 3 to 
 4.5\,m long  & \cite{comat} & $\sim$5000 \\\hline
 Aluminium Cover Foil & 10\,$\mu$m thick Al-foil & \cite{alfol} & 
 $\sim$6000\,m \\\hline
 \hline
 \end{tabular}\\[1mm]
 \end{center}
 \label{tab:material}
\end{table}

\paragraph{Cathode Foil:} 
The cathode foils form the honeycomb structure of the drift cells 
and ensure the mechanical stability of the modules (see cross sections in 
Fig.\ \ref{fig:single_doublelayer}).
To minimize secondary interactions in the detector, 
the cell walls must be as thin as possible. 
A conductive foil type developed by Bayer \cite{bayer}, 
commercially available from the company LONZA \cite{lonza}, 
has the required properties which 
are summarized in Table\ \ref{tab:foil}.
However, prototype modules built from this foil showed 
severe aging when operated at HERA \cite{otr_aging}.
This was caused by insufficient surface conductivity 
of the foil \cite{otr_aging2}. 
After detailed investigations, the problem could be solved
by metal-coating of the foil. 
Using a plasma coating process, the foils were covered 
with a 50\,nm thick copper and a 40\,nm thick gold layer \cite{apvv}. 
Copper alone is corroded by the fluoric radicals 
created by gas amplification in the Ar/CF$_4$/CO$_2$ drift gas mixture, 
and a direct gold coating has not 
sufficient adhesion on the Pokalon-C surface.

\begin{table}[htb]
 \caption{Properties of the cathode foil.}
 \begin{center}
   \begin{tabular}{|c|c|}
   \hline\hline
   Product name     & Pokalon-C                      \\
   Material         & polycarbonate cast film with 6 $\%$ soot \\
                    & (82\,$\%$ C, 13\,$\%$ O, 4\,$\%$ H, $<$1 $\%$ Cl) \\
   Residual solvent & $<$ 1.5\,$\%$                   \\
   Thickness        & 75\,$\mu$m                      \\
   Density          & 1.35\,$\textrm{g/cm}^{3}$       \\
   Tensile strength & 3800\,$\textrm{N/mm}^{2}$       \\
   Thermal expansion & 0.07\,mm/m/K                   \\
   Water vapour permeability & 14\,$\textrm{g/m}^2/\textrm{day}$         \\ 
   Additional coating & copper (50\,nm) / gold (40\,nm) \\
   Surface resistance & $<$ 1\,$\Omega$/$\square$        \\
   \hline\hline
   \end{tabular}
 \end{center}
 \label{tab:foil}
\end{table}

The first step in foil mass production was the folding of about 
12000 foils of 1\,m length and 40\,cm width.   
The device and the folding procedure are described in detail in
\cite{nikhef_hcsc}. 
The folded foils are then tempered between two aluminum templates 
for about 20 minutes at a temperature of 120$^{\circ}$\,C. In this
way an accurate hexagonal shape with straight cell sides is obtained. 
The last step is the copper and gold coating of the foil surfaces.

\paragraph{Base Plates:}
The base plates are glued to both module ends (see 
Fig.\ \ref{fig:mod_prod}). They have accurate holes for the
positioning and slits for the fixation of the module in the 
superlayer frame.  
The material is an epoxy/glass fibre multilayer (FR4) with very good 
mechanical properties and excellent chemical resistance.  
The plates are 70\,mm wide to allow the mounting of high voltage boards,
which provide the anode wires with high voltage and transfer the anode 
signals via twisted pair cables to the amplifier-shaper-discriminator 
(ASD-8) boards located outside of the gas box \cite{otr_fee}. 
In order to reduce the amount of material, 
no base plates are used at the inner end of modules above and below 
the proton beam pipe. 

\paragraph{End-Pieces:}
The end-pieces (see Fig.\ \ref{fig:mod_prod}) are plastic parts 
produced by a mould 
injection technique. In the first cell layer the end-pieces are glued  
onto the foil and the base plate. 
In the following layers, they are glued to each other and 
onto the foils. This provides a stable cell structure 
and a solid connection between the cells and the supporting
part of the module. The end-pieces provide openings for the gas 
flow through the cells and for the connection of the anode wires 
with the high voltage boards.

\begin{figure}
   \epsfig{file=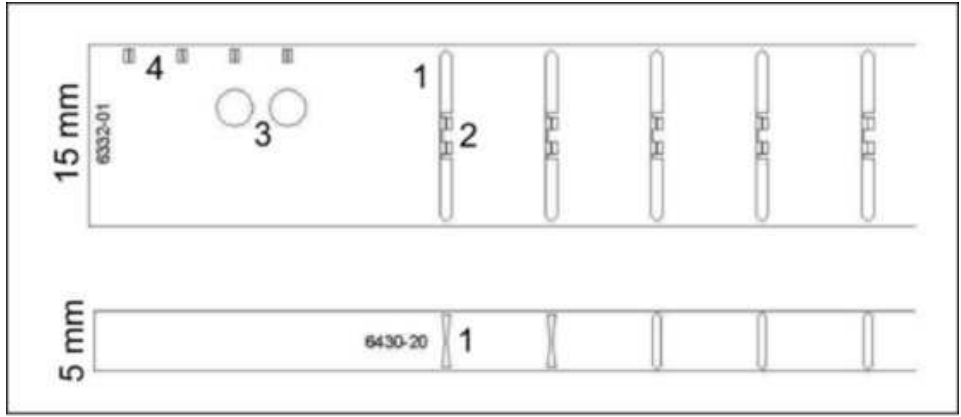,width=\linewidth}
   \caption{Left part of the two types of wire support strips.
     1: soldering pad, 2: pad for wire positioning, 3: holes for
     strip fixation, 4: pads for strip alignment.}
   \label{fig:strips}
\end{figure}

\paragraph{Wire Support Strips:}
The wires are kept in the centers of the hexagonal drift cells by
means of wire support strips. As shown in Fig.\ \ref{fig:strips},
two types of strips are used, both made of 100\,$\mu$m thick FR4 
laminate. The broad strips are used to exactly position the wires, 
to fix them by soldering, and to connect them to the external 
circuitry. These strips are glued onto the foil at both module ends 
(see Fig.\ \ref{fig:mod_prod}). 
For modules with segmented wires they are also used at the inner 
sector borders (see Fig.\ \ref{fig:segmentation}). 
On both ends they have precision holes for the fixation and 
pads with 50\,$\mu$m slits for the alignment 
relative to an external reference wire. The gold-plated copper pads 
for the anode wire positioning and fixation have a 
complicated structure. There are short parts for the wire positioning 
in 50\,$\mu$m slits and long parts for soldering, one for the
anode wire and one for either the connection to the HV board or for
the corresponding thick wire in an insensitive module section 
(see Fig.\ \ref{fig:segmentation}).
All four parts are connected electrically by thin bridges 
which prevent heat transfer when the second wire is soldered. 

In order to prevent electrostatical instabilities due to slight module
deformations, the free wire length must not exceed 60\,cm. In cases
where the distance between the broad strips is larger than that, an
appropriate number of narrow support strips is equidistantly inserted. 

\paragraph{Glue:}
Epoxy is used to glue the foils to each other, to glue the wire 
support strips and to glue the end-pieces to the foils.
As the outgassing of the epoxy is a potential source of chamber
aging, Stycast\,1266A, an epoxy with very low outgassing is chosen 
\cite{stycast}.

To improve the electrical contact between the foil layers, dots of
conductive silver-loaded epoxy are distributed every 30\,cm along 
the cell structure.

\paragraph{Wires:}
The anode wires are gold-plated tungsten wires of 25\,$\mu$m
diameter. The wire tension is 50$\pm$5\,g, and the 
elasticity limit is 130\,g.

Thick copper/beryllium wires of 75\,$\mu$m diameter are used for 
the signal propagation from sensitive inner module sectors to the 
high voltage board and the signal read-out at the module end
(see Fig.\ \ref{fig:segmentation}).
The large wire diameter prevents significant gas amplification 
at the nominal operating voltages. 
Both wire types are connected by soldering them 
to the wire support strip.
The thick wires are strung with the same tension as the thin ones to 
balance the forces on the strips.

\paragraph{Solder Tin:} 
A halogen-free solder tin with a reduced solder-forming flux is 
used to avoid larger amounts of flux residuals around the solder 
point on the wire support strips. 
A mask positioned around the strip protects
the cathode foil during soldering. The temperature of the 
soldering iron is required to be 280$^{\circ}$\,C. 

\paragraph{Carbon Fiber Rods:} 
Carbon fiber rods of 2\,mm diameter are used to increase 
the stiffness of modules longer than about 2\,m. 
Three rods are glued to both external module sides. 

\paragraph{Aluminum Cover Foils:} 
A further improvement of the mechanical stability is reached by covering 
both module sides with aluminum foils of 10\,$\mu$m thickness.

\subsection{Production Tools}
%%===========================
\label{sec:subsectPrTools}

In order to produce stiff honeycomb drift tube modules with precisely 
defined dimensions from foils which by themselves have no 
stability at all, the foils are handled using templates which force
them to assume the nominal cell shape during assembly.
Since the module length varies from 75\,cm to 446\,cm, 
a modular template system is designed. The complex parts at both
ends and in the center, where the strips for the wire positioning 
and fixation are mounted, are machined from aluminum
plates. The simpler parts in between are made of metal-reinforced 
epoxy using a casting technology. 
They have high precision and form stability but at much 
lower cost. 
In the template grooves there are holes through which under-pressure 
is applied which safely draws the foils into shape.

\begin{figure}
   \epsfig{file=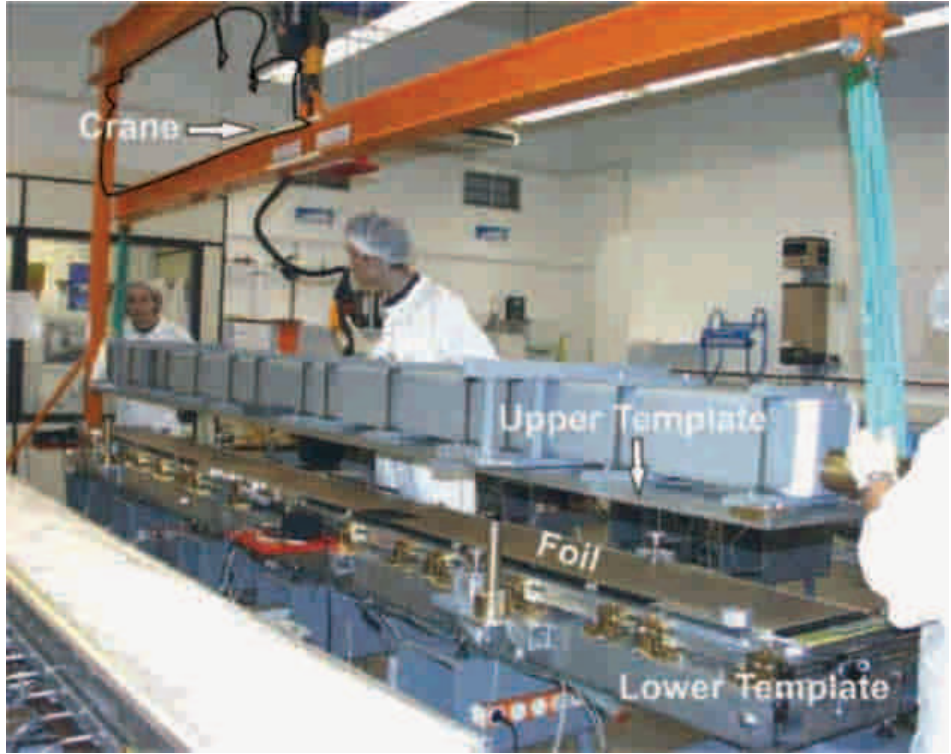,height=10cm}
   \caption{Production workplace with the upper template  positioned 
     above the lower one just before closing the next cell layer.}
   \label{fig:template1}
\end{figure}

Each module production workplace consists of an upper and a lower
template (Fig.\ \ref{fig:template1}). After placing the first set 
of foils, the wiring as well as wire tension and high voltage 
stability tests are performed on the open cells in the lower template.
The upper template carries the set of top foils to be glued. 
The glue is distributed with a foam roller on the enhanced part 
of the foil profile in the upper template.
A special mask made of a thin steel foil guarantees
that the glue covers only the horizontal foil area and not the inner cell 
walls. 
Then the upper template is lifted by a crane, rotated and positioned 
above the lower template.
With the help of positioning devices it is placed precisely onto the lower 
one thus closing a monolayer.  

For each working place the template parts are assembled and aligned 
on a stable frame. 
Since the upper template has to be lifted and rotated,
a movable crane is used for the long templates of the PC and TC modules. 
The template tolerances for the cell geometry and for the distance from 
the first to the last cell are better than 80\,$\mu$m. The alignment 
accuracy of the template parts is about 50\,$\mu$m.

\subsection{Module Production}
%%======================================

The main assembling steps are demonstrated in 
Fig.\ \ref{fig:mod_prod} for the first monolayer of a module.
The module production starts with positioning the lower foil 
and the base plates in the lower production template. 
On both sides the end-pieces are glued to the foil and to the 
base plate giving a stable connection between both.
In the first cell monolayer, wire supporting strips are glued
starting 7\,mm from the foil ends.
The next steps are the wiring of the monolayer, the test of the wire 
tension and of the high voltage stability. 
If necessary, bad wires are replaced.
Then, epoxy is distributed on the foil in the upper template and
the layer is closed. 
After the curing of the glue, which lasts about three hours at room 
temperature, the procedure is repeated for the next monolayer.
Finally, the carbon fiber rods and the aluminum cover foil are
glued onto both module sides, and the high voltage
boards are mounted. 

To achieve the required throughput, the production was organized in different 
laboratories 
(NIKHEF Amsterdam, IHEP Beijing, Tsinghua Univ. Beijing, 
JINR Dubna, DESY in Hamburg and in Zeuthen). 
For the mass production standardized production templates, tools and 
the same detector materials were distributed to all institutes. 
In addition, standardized production procedures and quality tests 
have enforced coherent production conditions.
In total 22 workplaces were installed in the clean rooms of the 
collaborating institutions, 17 of them for the production of  
PC- and TC-modules, and the others for the MC-modules.

With two shifts per day, a standard module with 4 monolayers can be built 
and tested within three days. The limiting factor is the 
curing time of the glue, but at each institute there were enough
workplaces to avoid idle time. The production of the 978 modules 
could be finished within 15 months. 

\subsection{Quality Tests}
%%=============================

The nominal wire tension is 50\,$\pm$\,5\,g 
and wires outside of the tolerance limit have to be replaced. 
Before the monolayer is closed by the upper foil, the average 
dark current of all wires is measured in air using a high 
voltage of 2000\,V. 
The allowed maximum current is 30\,nA at low 
air humidity. Wires with larger currents are 
cleaned and have to be replaced if the current is still above 
the limit.

Each module production site operates a test stand for high 
voltage training and single wire test of the produced modules. 
The modules 
to be tested are put into a gas-tight box which is then flushed 
with an Ar/CO$_2$ (50:50) gas mixture. The high voltage is ramped 
up to the nominal values of 1850 (2230)\,V for the 5 (10)\,mm drift 
cells. Only in rare cases the dark currents of single HV groups 
are so high that this has to be done in steps. Wires which after 
12 hours at nominal HV still show dark currents $>1\,\mu$A are 
considered unusable and are disconnected.

After the training, the analog signals from all wires are checked.
This is done using either a 
cosmic particle trigger setup or a radioactive source together 
with an oscilloscope and allows to identify dead and noisy wires. 
Of the 112\,674 wires in the Outer Tracker 482 are 
disconnected because of shorts or excessive dark currents, 363
showed no signals, and 402 have noise rates $> 100$\,kHz. 
This corresponds to a fraction of 1.1\,$\%$ bad wires.

The results of quality control during module production and 
testing are recorded in a document file which accompanies each 
module from the start of production till the installation in a 
chamber. A computer readable test sheet per module can be 
inspected via a graphical web interface and is used during data 
taking for data quality cross-checks.

%%% Local Variables: 
%%% mode: latex
%%% TeX-master: t
%%% End: 

\section{Superlayer Structure and Assembly}

\begin{figure}
   \epsfig{file=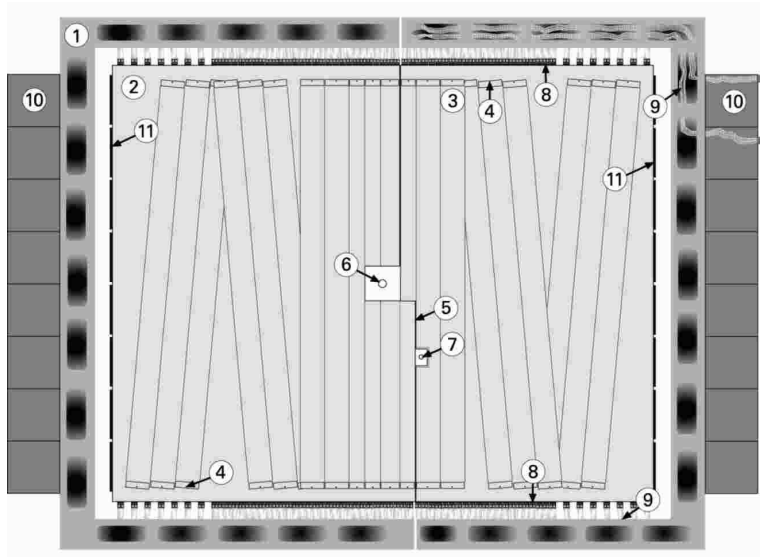,width=10cm}
   \caption{Schematical view of the superlayer structure.
   1 - outer steel frame, 2 - sandwich panel closing the
   gas box, 3 - drift cell modules, 4 - module fixation pin, 
   5 - overlap region of superlayer halves, 6 - proton beam pipe,
   7 - electron beam pipe, 8 - ASD-8 boards, 9 - signal cables to 
   TDC, 10 - TDC crates, 11 - low voltage distribution boards.}
   \label{fig:sl_schemat}
\end{figure}

Table\ \ref{tab:otr_sl} summarizes the
positions, dimensions and stereo layer configurations of all  
Outer Tracker superlayers.
A schematical view of the structure which is typical for PC and
TC superlayers is given in 
Fig.\ \ref{fig:sl_schemat}. It shows the division into two independent
chambers across the proton and electron beam pipes.
In their nominal position both chambers overlap to avoid
detection inefficiencies for track reconstruction. 
The chambers are 
mounted on rails which allows for a movement perpendicular 
to the beam axis for easier installation and repair. 
Each chamber consists of two main components, 
the gas box and the outer steel frame.

\subsection{Gas Box with Detector Modules} 

The gas box of a chamber fulfills several requirements: 
\begin{itemize}
\item It allows for a precise mounting of the modules in all 
  stereo layers.
\item It provides a gas tight enclosure for the modules which is
  filled with the counting gas.
\item It places as little material as possible in the 
  acceptance area of the detector.  
\end{itemize}
The box consists of a C-shaped aluminum frame, a thin carbon 
fiber cap which closes the C and covers the modules in the
overlap area, and two large sandwich 
panels closing the front and the back of the box.

\begin{figure}
\begin{center}
   \epsfig{file=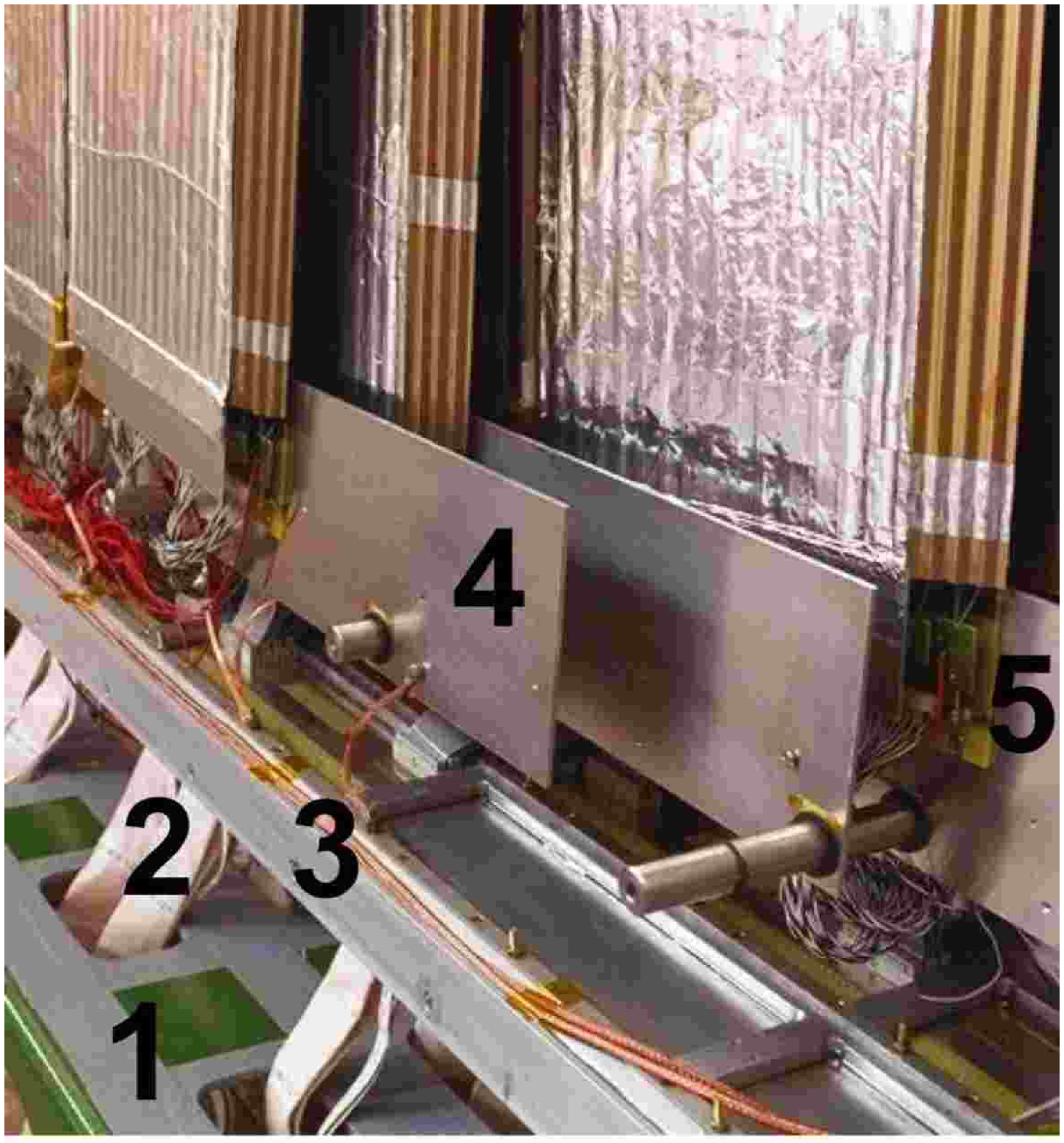,height=7.5cm} \hspace{0.2cm}
   \epsfig{file=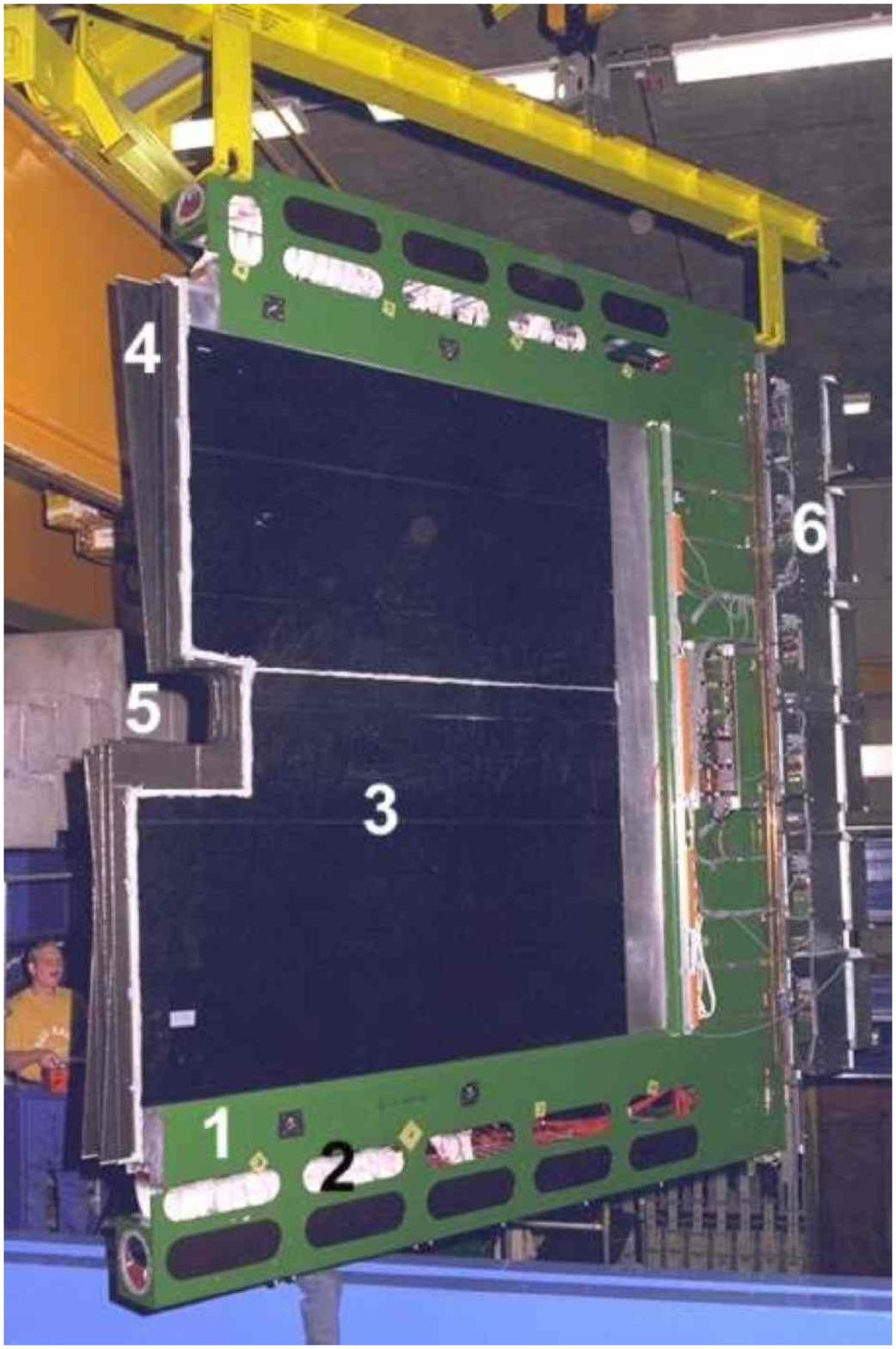,height=7.5cm}
   \caption{
   Left picture: Lower part of the TC1 chamber. 
   1: outer steel frame, 2: signal cables which are 
   plugged on the ASD-8 boards (not visible), mounted at the 
   lower side of the gas box frame, 3: gas box frame,
   4: plates for module fixation and a part 
   of the installed modules in the three stereo layers, 5: base plate 
   of a module with high voltage board.
   Right picture: Completed PC2 chamber. 1: outer steel frame, 
   2: signal cables, 3: large sandwich panel closing the gas box, 
   4: structured end-cap for the overlap of modules from both 
   chambers, 5: cut-out for the proton beam pipe,
   6: TDC crates.}
   \label{fig:tc_close}
\end{center}
\end{figure}

The picture of an open TC chamber in Fig.\ \ref{fig:tc_close} 
shows the gas box frame together with plates on which the modules are 
mounted. By means of snap rings the plates are held in well-defined 
positions on large dowel pins which are anchored in the frame. 
This defines the z-positions of the stereo layers. 
On the plates there are many holes for pins and screws which are 
needed for vertical and horizontal module positioning and fixation.

The metal frame of the gas box has cut-outs into which so-called
feed-through boards are inserted. These provide gas-tight connections
of signal and HV lines from the modules inside to the outside of the 
gas box. On the outside, the ASD-8 boards are directly plugged onto
the feed-through boards.

The large plates which close the gas box have to withstand large 
forces even at only 0.5\,mbar overpressure inside the gas volume.
On the other hand, the plates have to consist of light-weight
material to minimize their contribution to the total material
budget in the detector acceptance area.
The plates are sandwich panels with an aramid paper core of 
hexagonal cell structure covered by two plates of carbon or 
glas fiber epoxy composite for the PC or TC chambers, respectively 
\cite{hexel}. 
The cover plates with dimensions of about 2.5 $\times$ 3.2 
$\textrm{m}^{2}$ for PC and 3.5 $\times$ 5 $\textrm{m}^{2}$ 
for TC chambers are glued on the gas box
frame and additionally fixed by screws. The inner side of the 
plates is covered by a thin aluminum foil which provides an
electrical shielding and prevents the contamination of the 
drift gas by outgassing products of the plate epoxy. 

In the overlap area of both chambers 
end-caps made of a 1\,mm thick carbon fiber epoxy composite
material close the gas boxes. The right picture in
Fig.\ \ref{fig:tc_close} shows the cap of the PC2 chamber.
Its structure is rather complicated since the cap has 
to enclose modules of six different stereo layers, to assure 
appropriate overlap and to follow the cut-out for the beam pipe. 
The caps were produced using templates 
which could be adjusted to the lengths and layer configurations of all 
chambers \cite{LH}.

\subsection{Outer Steel Frame} 

The outer steel frame is designed such that it can carry the 
weight of the gas box and the cables. The weight of a complete 
TC chamber is about three tons.
The C-shaped steel frame is open towards 
the  other chamber. Within the frame, the gas box 
containing the detector modules is mounted. 
The rectangular profile of the frame houses the large amount 
of signal cables, which are twisted pair cables connecting the ASD-8 
front-end boards on top and bottom of the gas box with 
the TDC boards installed in crates at the vertical 
outer side of the frame. 
The holes in the steel profile seen in the left picture of 
Fig.\ \ref{fig:tc_close} facilitate the insertion of
cable bundles.

\subsection{Assembly and Installation Steps}

The chamber assembly starts with routing of prepared signal cable 
bundles in the outer frame from the position of the feedthrough 
boards to the corresponding TDC crates. Next, the
gas box, still without front cover plate and overlap caps, is 
installed in the cabled frame. For module installation the chamber 
is brought to an upright position in a large dust-protection tent,
in order to avoid extreme bending of the modules during 
installation.

The modules are installed layer by layer. After bringing all 
module fixation plates of a layer into their nominal positions, 
the modules are positioned and screwed on the top plates. 
On the bottom plates only the lateral position is fixed by a dowel 
pin in a precise slit in the module base plate, 
thus allowing for thermal expansion of the modules without 
buckling. Then, the signal cables are plugged and the high voltage 
cables are plugged soldered to the corresponding connectors on the 
inner sides of the feedthrough boards. 
Only after a careful check of all electrical 
connections and of the high voltage stability of the just installed 
modules the chamber assembly proceeds to the next stereo layer.

After all modules are installed, the front cover plate is glued 
and screwed to the gas box frame. The gas box is closed by mounting 
the cap structure in the overlap region with the other chamber. 
After checking the gas tightness of the assembled chamber it is 
equipped with ASD-8 amplifier boards on the feedthrough boards and 
with TDC boards in the crates on the outer frame. The low and high 
voltage distributions are installed as well. Hence all electronics 
can be commissioned using test pulses before the chamber is
installed in the experiment.

Special support structures are used for transporting the chambers 
from the assembly to the experimental hall. 
In the HERA-B detector there is a rail for each superlayer 
from which both chambers are suspended. 
This allows chamber movement to and from the beam pipe which facilitates
installation and maintenance of the Outer Tracker. 
Even for the largest chambers, the installation with the connection 
of the external cables and gas pipes takes no longer than a day.

\subsection{Description of the Detector Geometry}

In order to achieve a maximum of detector hermeticity with 
a minimum amount of material, the OTR detector geometry became 
rather complicated. 
The design of the superlayers and of the different module types 
was performed using a CAD system. The information on module 
and superlayer types as well as their geometrical description 
is stored in a data base. 
For the Monte Carlo simulation with GEANT \cite{geant},
in addition to the geometrical data, all material properties 
of gas boxes and detector modules are taken into account.
In total about 2500 different detector volumes are described,
which are the module sectors (see Fig.\ \ref{fig:cell_segm}). 

\begin{figure}
\begin{center}
   \epsfig{file=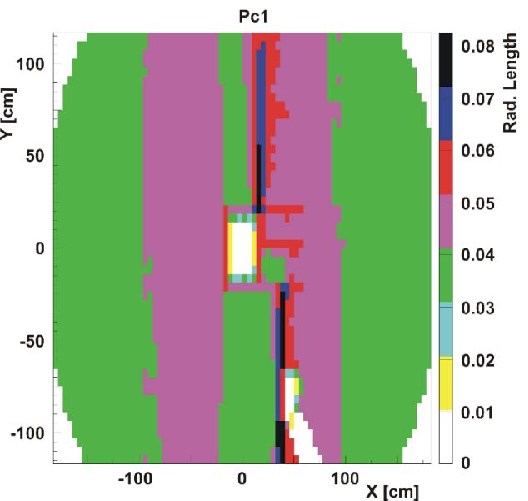,width=5.5cm} \hspace{0.5cm}
   \epsfig{file=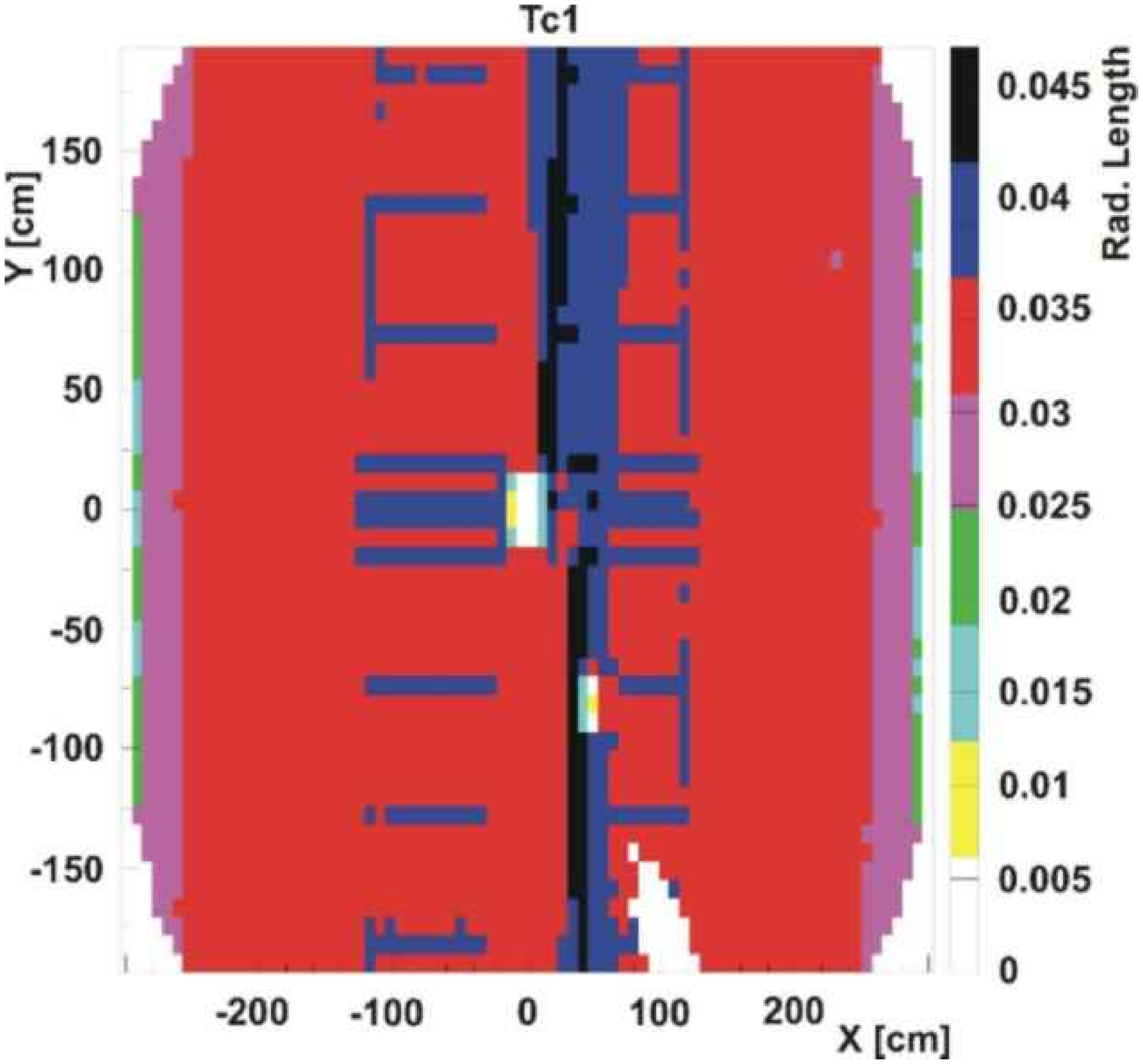,width=5.5cm}
   \caption{Thickness of superlayers PC1 (left) and TC1
     (right) in radiation lengths.}
   \label{fig:radlen}
\end{center}
\end{figure}

In Fig.\ \ref{fig:radlen} the material distribution is shown 
in terms of radiation lengths for the PC1 and
TC1 superlayers. On average PC1 has 4--5\,\% of a 
radiation length, TC1 only 3--4\,\%.

Most of the material is concentrated in the overlap region 
of both chambers of each superlayer. 
In the area of TC1 for the 5 mm modules the wire support
strips are visible. 
These areas are rather broad due to staggering of the strips  
within a module and the inclination of the stereo layers.
Summing up the material contributions of all 13 superlayers,
one gets about 0.4 radiation lengths on average and about 
0.7 radiation lengths in some overlap areas of the chambers.

With the changed physics programme of HERA-B for the 2002 run the
reconstruction efficiency for $K^0_S$ decays was of minor importance.
It was found to be more crucial to reduce the background in the
ECAL from secondary interactions in the magnet. As a consequence,
the superlayers MC2--MC8 were removed from the detector which reduced
the average material contribution of the Outer Tracker to 
0.3 radiation lengths.

%%-------------------------------------------
%%\marginpar{\tiny
%%TOTAL OTR 
%%THICKNESS,
%%MIN, MAX, AVER.
%%}
%%-------------------------------------------

%%% Local Variables: 
%%% mode: latex
%%% TeX-master: "commission.txt"
%%% End: 

\section{Gas System}

The drift velocity of the counting gas has to be large because of 
the 96\,ns time distance between bunches and the requirement 
to register events within this time. 
With a gas mixture containing a sufficient
fraction of the fast gas component CF$_4$, 
drift velocities of up to 150\,$\mu$m/ns are reached \cite{grimm}. 
After studies of aging and etching properties with 
different gas mixtures like CF$_4$/CH$_4$, Ar/CF$_4$/CH$_4$ and
Ar/CF$_4$/CO$_2$ \cite {otr_aging}, the latter mixture was 
chosen with a volume ratio of 65\,$\%$/30\,$\%$/5\,$\%$. 

\begin{figure}
   \epsfig{file=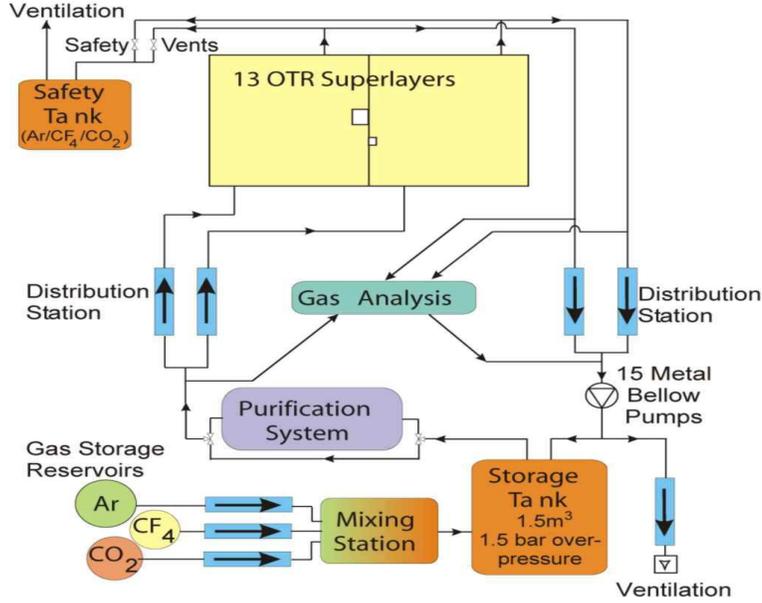,width=10cm,
     height=8cm}
   \caption{Schematical view of the OTR gas system.}
   \label{fig:gas_scheme}
\end{figure}

The total volume of all OTR chambers is about 22\,m$^3$.
Because of the high cost of CF$_4$ and also for environmental 
reasons, it is necessary to use a closed-loop gas system. 
The block diagram in Fig.\ \ref{fig:gas_scheme} shows the 
functionality.
The circuit consists of the following components: 
A 1.5\,m$^3$ storage tank where fresh gas coming from 
the mixing station is added to the circulation, the purification system, 
the distribution station which regulates the gas flow into
the 26 chambers, and the pump station with 15 metal bellow
pumps which circulate the gas.
Both, gas flow and pressure are regulated and controlled with help of 
a programmable logic controller for the 
pressure transducers and mass flow controllers.
The flow rate through the system is 20\,m$^3$/h, i.e. almost 
one volume exchange per hour.
The gas enters a chamber through a distribution pipe at the bottom, 
flows through the cells of the modules, and leaves the gas box at 
the top. In order to minimize air intake through leaks in the gas boxes, 
the pressure in all chambers is kept at 0.5\,mbar above atmospheric
pressure.
A safety tank filled with counting gas is acting as a buffer volume
to minimize air intake in an alarm situation 
when the regulation is switched off and the chambers open to the 
outside via safety valves and safety tank.
A detailed description of the gas system and its performance is 
given in \cite{hohlmann}.
 
\paragraph{Gas Storage and Mixing:} 
There are two 450-liter gas storage reservoirs filled with 
liquid Argon and with CO$_2$, respectively, and a 600-liter 
tank for CF$_4$. 
The gases flow into the mixing station in ratios
controlled by gas flow regulators and then into the storage tank.
In case of the first fill or for a longer shut-down period,  
a cheaper Ar/CO$_2$ mixture is used to remove the air or 
to prevent that air enters the closed-loop system.
With addition of the CF$_4$ component one obtains the proper 
working gas mixture within several days.
During data taking about 0.5\,to\,2\,$\%$ fresh gas per volume
exchange is continuously added to the circulating gas.
This results in a CF$_4$ consumption of about 400\,kg/month.
The required tolerances for the gas composition are 65$\pm$1$\%$, 
30$\pm$1$\%$ and 5$\pm$0.2$\%$ for Argon, CF$_4$ and CO$_2$, 
respectively.

\paragraph{Purification Station:}
Before the gas is distributed to the different chambers, it
is purified by removing oxygen and water. 
There are two regenerable purifier stations installed, both 
containing ~60\,kg of catalyst R3-11G \cite{purifier}. While one of 
them is in operation, the other station is excluded 
from the closed-loop circuit for regeneration. 
About every 10 days the stations are switched. 
This guarantees a continuous gas purification keeping the oxygen 
content below 200\,ppm and the water level below 20\,ppm.  
Mainly due to residual air in the system and because of 
small leaks, there is a non-negligible fraction of nitrogen 
in the counting gas. 
The nitrogen level is kept below 2000\,ppm by continuously venting 
a small fraction of the circulating gas and by adding a corresponding 
amount of fresh gas. 

Trace contaminants in the counting gas were suppressed by strictly
avoiding the use of outgassing materials in fittings, seals, pumps
etc. Stainless steel and flexible metal tube piping was used
throughout the system.

\paragraph{Gas Analysis for Quality Control:}
The system continuously analyses the gas by measuring the fractions 
of the three main components and of certain impurities like 
oxygen, nitrogen and water. The common input after purification 
and all output lines from the 26 chambers can be connected
via a programmable switch to the gas analysis station which 
consists of a gas chromatograph, an oxygen meter and a moisture
meter. Each measurement takes about 10 minutes, i.e. after 4.5 hours 
a given line is measured again.

%%% Local Variables: 
%%% mode: latex
%%% TeX-master: "commission.txt"
%%% End: 

\section{Summary}

The Outer Tracker of the HERA-B experiment provides 90\,\% solid angle
coverage in the center-of-mass system and is capable of recording up
to 200 charged particle tracks every 96\,ns. 

In total 13 superlayers, each
consisting of two individual planar drift chambers, were assembled and
installed in three areas of the experiment: 7 inside the magnet,
4 between magnet and RICH, and 2 in front of the electromagnetic
calorimeter. The stereo layers inside each chamber are composed of
honeycomb drift tube modules. Module assembly from folded Pokalon-C 
foils results in a close-packed hexagonal drift cell structure. 
The technique of installing anode wires into open drift cells 
facilitates also longitudinal
wire segmentation which is needed for limiting the occupancy close to 
the proton beam. Chamber aging, observed with prototype modules at 
high hadron fluxes, was cured by coating the cathode foils with thin 
layers of copper and gold, and by a proper drift gas. 
In total 978 detector modules with about 
113000 electronics channels have been built. Two thirds of the modules
have 5\,mm diameter drift cells, the cell size in the others is 10\,mm.
Over a period of 15 months module production proceeded in parallel at 
six different laboratories, using the same tools, materials, and 
instructions. The initial fraction of bad channels was 1.1\,\%.

The honeycomb drift tube modules offer clear advantages in terms of
operational safety, cathode aging, and easy mechanical handling.
Although made of light material, the total amount of scattering
material was on average about 0.4 radiation lengths in the fiducial 
volume of the detector. For the 2002 run this
was reduced to 0.3 radiation lengths by removing chambers from the magnet.

For the 26 chambers of the Outer Tracker, with a total volume of 
22\,m$^3$, a closed-loop gas system was built. It allows 
to regulate the flow, to control the pressure of the 
Ar/CF$_4$/CO$_2$ (65:30:5) gas mixture in all chambers, and to 
purify the gas.

During its operation from December 2001 until March 2003, the Outer
Tracker of HERA-B worked well, as will be described in more detail in 
a forthcoming paper on OTR performance \cite{perform}. 

In conclusion, it has been
shown that a large tracker can be efficiently built and safely operated
under high radiation load at a hadron collider.

\section*{Acknowledgements}
%\begin{ack}
We thank our colleagues of the HERA-B Collaboration who made in a
common effort the running of the detector possible. The HERA-B
experiment would not have been possible without the enormous effort
and commitment of our technical and administrative staff. It is a
pleasure to thank all the teams at different sites involved in 
prototype chamber development, in module design and production, 
in construction and assembling of superlayers, and in the 
design and realization of the gas system.

We express our gratitude to the DESY laboratory for the strong support
in setting up and running the HERA-B experiment. We are also indebted
to the DESY accelerator group for the continuous efforts to provide
good beam conditions.

\vspace{2cm}
%\end{ack}

%\end{runninglinenumbers}

\end{document}